\documentclass{aa}  
\usepackage{natbib}
\bibpunct{(}{)}{;}{a}{}{,} 

\usepackage[T1]{fontenc}
\usepackage{graphicx}
\usepackage{amssymb}
\usepackage{subfigure}
\usepackage{epstopdf}
\usepackage{gensymb}
\usepackage{siunitx}

\usepackage{soul}

\usepackage{mathtools}
\usepackage{commath}
\usepackage{diagbox}
\usepackage{makecell}
\usepackage{tabularx}
\usepackage[final]{changes} 
\usepackage{cancel}

\begin{document}

   \title{System Equivalent Flux Density of a Polarimetric Tripole Radio Interferometer}


   \author{A. T. Sutinjo
          \inst{1}
           \and
           M. Kovaleva
          \inst{1}
          \and
          Y. Xu
          \inst{2}
          }

   \institute{International Centre for Radio Astronomy Research (ICRAR), Curtin University, 6102 Australia\\
              \email{adrian.sutinjo@curtin.edu.au}
       \and
       National Astronomical Observatories,  Chinese Academy of Sciences, Beijing 100101, China \\
             }

   \date{
   Accepted by PASP, \today
   }

 
  \abstract
  {Electromagnetic and information properties of tripole antennas have been studied since the 1980s. In radio astronomy, tripole antennas find an application in space telescopes. More recently,  a radio interferometer with satellite-borne tripole antennas is being considered for a lunar orbiting radio telescope to observe very long wavelengths.  } 
   {System equivalent flux density (SEFD) is an important figure of merit of a radio telescope. This paper aims to derive a general expression for SEFD of a polarimetric tripole interferometer. The derivation makes only two basic and reasonable assumptions. First, the noise under consideration is zero mean and when expressed in complex phasor domain, has independent and identically distributed (iid) real and imaginary components. Correlated and non-identically distributed noise sources are allowed as long as the real and imaginary components remain iid. Second, the system noise is uncorrelated between the elements separated by a baseline distance. }
   {The SEFD expression is derived from first principles, that is the standard deviation of the noisy flux estimate in a target direction due to system noise.}
   {The resulting SEFD expression is expressed as a simple matrix operation that involves a mixture of the system temperatures of each antenna and the Jones matrix elements. It is not limited to tripoles, but rather, fully extensible to multipole antennas; it is not limited to mutually orthogonal antennas. To illustrate the usefulness of the expression and how the formula is applied, we discuss an example calculation based on a tripole interferometer on lunar orbit for ultra-long wavelengths observation. We compared the SEFD results based on a formula assuming short dipoles and the general expression. As expected, the SEFDs converge at the ultra-long wavelengths where the dipoles are well-approximated as short dipoles. The general SEFD expression can be applied to any multipole antenna systems with arbitrary shapes. }
   {}


   \maketitle
%
\section{Introduction}
\label{sec:intro}
It has been recognized since 1980s that tripole antennas have interesting electromagnetic properties that enable interference mitigation~\citep{Compton_1142690}, null steering~\citep{Compton_1143119} and direction finding~\citep{Ladreiter1995RaSc...30.1699L,Cecconi2004RS003070, Chen_5666797, Wong_937478}, to name a few applications. In planetary and solar science, we find tripole antennas as space probes mounted on satellites~\citep{Rucker96RS01972, Cecconi2004RS003070, Bale2008, ZARKA2012156, Fischer_2021RS007309}. More recently in radio astronomy, there have been increasing interests in tripole antennas as elements for radio astronomy imaging~\citep{Carozzi_10.1111/j.1365-2966.2009.14642.x}, moon-based radio astronomy~\citep{KLEINWOLT2012167}, and in particular, for a lunar orbiting radio interferometer~\citep{Boonstra_7500678, Bentum_2020AdSpR..65..856B, Chen_2018ExA....45..231C, Chen_doi:10.1098/rsta.2019.0566, Huang_2018, Rajan2016ExA....41..271R,Arts_8739399}. Space missions are costly, and hence a space mission plan requires a reliable model-based prediction of key system performance such as sensitivity. 


System equivalent flux density (SEFD) is a figure of merit that characterizes the sensitivity of a radio interferometer~\citep{1989ASPC....6..139C, 1999ASPC..180..171W}. It expresses the system noise due to the instrument and the sky under observation as an equivalent flux density of an incident wave from a target direction. For a dual-polarized antenna system~\citep{1989ASPC....6..139C, 1999ASPC..180..171W}
\begin{eqnarray}
\text{SEFD}\approx\frac{1}{2}\sqrt{\text{SEFD}_{XX}^2+\text{SEFD}_{YY}^2},
\label{eqn:SEFD_XYapprox}
\end{eqnarray}
where $\text{SEFD}_{XX}, \text{SEFD}_{YY}$ are the SEFD of each antenna polarization assuming an unpolarized source; the approximation is valid under certain conditions~\citep[see][Sec.~4]{Sutinjo_AA2021}. The question for a tripole or multipole antenna system is how to properly extend the SEFD expression to these cases. The SEFD expression should not be derived based on an ad-hoc amendment to Eq.~\eqref{eqn:SEFD_XYapprox}, but rather with clarity regarding the most fundamental assumptions and with a guidance as to how to obtain the parameters in the formula.

This paper extends the treatment for the SEFD of a radio interferometer in~\citet{Sutinjo_AA2021}\footnote{As explained in this reference, our SEFD expression differs from conventional treatment that is based on antenna effective aperture over system temperature, $A_e/T_{sys}$, of a single polarized antenna. Rather, the new treatment is based on the noise of the flux estimate of the polarimetric intereferometer. For more extensive rationale and comparative literature review regarding this approach, we refer interested readers to sections~1 and~2 of this reference.} \added{(hereafter we refer to this paper as S2021)} from dual-polarized antennas to multiple-polarized antennas. Our particular focus, for illustration, is the tripole antenna. This is motivated by the lunar orbiting interferometer such as Discovering Sky at the Longest wavelength (DSL) array\footnote{The DSL array is also known as
the ``Hongmeng'' mission, which means ``Primordial Universe'' in Chinese.}~\citep{Chen_doi:10.1098/rsta.2019.0566}, consisting of a mother satellite and six to nine daughter satellites located on the same orbit around the Moon. Besides one daughter satellite dedicated to the high precision measurement of the sky-averaged spectrum, the other daughter satellites form a reconfigurable linear interferometer array with variable spacing. The target frequency range of the interferometer is 30~MHz and below. In this particular application, the field-of-view (FoV) is the entire visible sphere surrounding the interferometer. Therefore, we need an SEFD expression which is valid for the entire sphere and for any polarization of the sources therein. To the best of the authors' knowledge, there is no published  formulation of SEFD for a tripole (or multipole) polarimetric interferometer other than ones based on approximate reasoning~\cite[see][for example]{ZARKA2012156}. 

To derive a general expression for direction-dependent SEFD, we take similar steps as~S2021 with two key differences. First, we derived the generalized estimate of the mutual coherence of electric field for the overdetermined system  that arises due to the tripole (or multipole) antenna. This is discussed in Sec.~\ref{sec:back}. Second, we produced a general expression of SEFD for the multipole antennas. The generality arose from a careful review of the assumptions in the statistical calculations for the standard deviation of the flux density estimate. We found that neither the assumption of unpolarized sky noise nor of orthogonal antennas is necessary. It is ample for our purpose to let the noise represented in the complex phasor domain have independent and identically distributed (iid) real and imaginary components with zero mean. This is fully consistent with naturally-occurring noise sources in radio astronomy as well as instrumentation noise such as that of the low noise amplifiers (LNAs). This discussion is summarized in the appendix of this paper and generalizes our prior treatment. The resulting general SEFD expression is given in Sec.~\ref{sec:Gen_SEFD}. A special case of the tripole antennas, envisaged for the DSL interferometer, is discussed in Sec.~\ref{sec:SEFD} and illustrated in Sec.~\ref{sec:Example} with a comparison to an electromagnetic simulation. Finally, concluding remarks are presented in Sec.~\ref{sec:concl}.

\section{Problem statement and basic approach}
\label{sec:back}

\begin{figure}[htb]
\begin{center}
\noindent
  \includegraphics[width=3.25in]{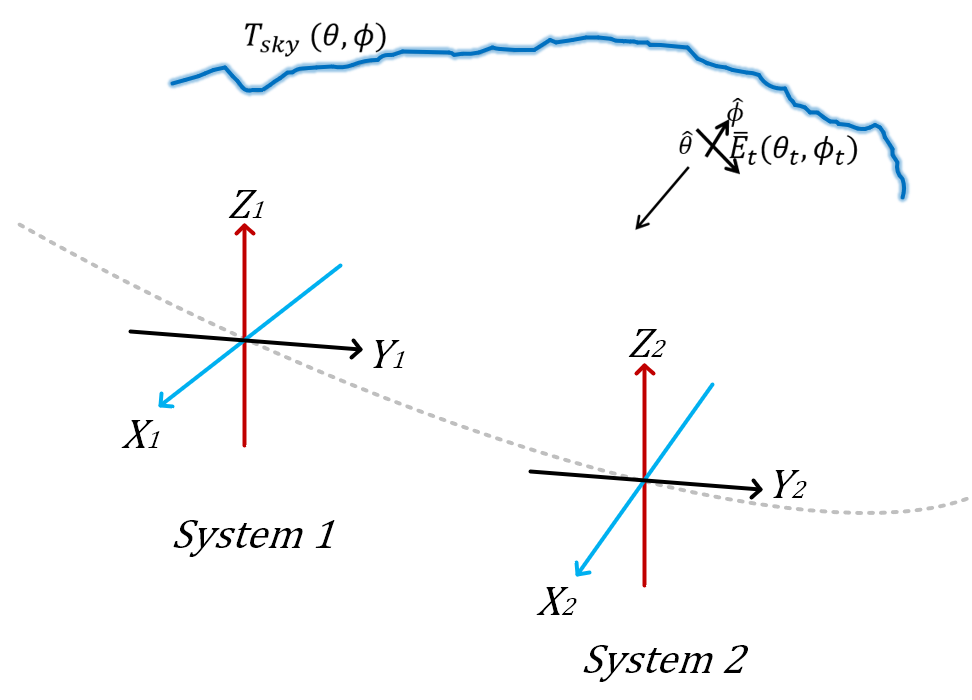}
\caption{Tripole interferometer under consideration. $T_{sky}$ is the distributed sky noise. $\bar{E}_t$ is the incident electric field from a target direction. Although a tripole system is shown, the idea extends to a multipole system.}
\label{fig:Tripole_sys}
\end{center}
\end{figure}

The problem under consideration is illustrated in Fig.~\ref{fig:Tripole_sys}. The voltages measured by the tripole antenna system~1 in response to the target electric field ($|_t$) are
\begin{eqnarray}
\mathbf{v}_1|_t&=&\mathbf{J}_1\mathbf{e}_t, \nonumber \\
\left[ \begin{array}{c}
V_{X1}|_t \\
V_{Y1}|_t \\
V_{Z1}|_t
\end{array} \right]
&=&
\left[ \begin{array}{cc}
l_{X1\theta} & l_{X1\phi} \\
l_{Y1\theta} & l_{Y1\phi} \\
l_{Z1\theta} & l_{Z1\phi} \\
\end{array} \right]
\left[ \begin{array}{c}
E_{t\theta} \\
E_{t\phi} 
\end{array} \right],
\label{eqn:J1}
\end{eqnarray} 
where $\mathbf{J}_1$ is the Jones matrix of system~1, with entries which represent the antenna effective lengths, with units of \si{\metre}~(see S2021); $E_{t\theta},E_{t\phi}$ are the components of the incident electric field vector, with units of \si{\volt\per\metre}. For the tripole, the Jones matrix has three rows, but the proposed approach is valid for any matrix with two or more rows. The voltages are correlated, resulting in the outer product 
\begin{eqnarray}
\mathbf{v}_1\mathbf{v}_2^H|_t&=&\mathbf{J}_1\mathbf{e}_t\mathbf{e}_t^H\mathbf{J}_2^H\nonumber \\
&=&
\left[ \begin{array}{ccc}
V_{X1}V_{X2}^*|_t & V_{X1}V_{Y2}^*|_t & V_{X1}V_{Z2}^*|_t\\
V_{Y1}V_{X2}^*|_t & V_{Y1}V_{Y2}^*|_t & V_{Y1}V_{Z2}^*|_t \\
V_{Z1}V_{X2}^*|_t & V_{Z1}V_{Y2}^*|_t & V_{Z1}V_{Z2}^*|_t 
\end{array} \right],
\label{eqn:outer_voltage}
\end{eqnarray}
where $\mathbf{J}_2$ is the Jones matrix of system 2, $.^H$ denotes conjugate transpose and $.^*$ indicates complex conjugation. The expected value of Eq.~\eqref{eqn:outer_voltage} is
\begin{eqnarray}
\left<\mathbf{v}_1\mathbf{v}_2^H\right>|_t=\mathbf{J}_1\left<\mathbf{e}_t\mathbf{e}_t^H\right>\mathbf{J}_2^H,
\label{eqn:exputer_voltage}
\end{eqnarray}
where $\left<.\right>$ indicates statistical expectation. The quantity of interest is 
\begin{eqnarray}
\left<\mathbf{e}_t\mathbf{e}_t^H\right>&=&
\left[ \begin{array}{cc}
\left<|E_{t\theta}|^2\right>& \left<E_{t\theta}E_{t\phi}^*\right> \\
\left<E_{t\theta}^*E_{t\phi} \right> & \left<|E_{t\phi}|^2\right>
\end{array} \right],
\label{eqn:outer_field}
\end{eqnarray}
from which the polarization property of the target may be inferred using the Stokes parameters $I, Q, U, V$~\citep{Wilson2009, Smirnov1_2011}. 

In practice, we obtain an estimate of Eq.~\eqref{eqn:outer_field} from the outer product  $\mathbf{v}_1\mathbf{v}_2^H$ which contains system noise. This is achieved through inversion of the Jones matrices. For a multipole system, however, the inverse of the Jones matrices do not exist. Nevertheless,  if the columns of the Jones matrices are made independent, $\mathbf{J}^H\mathbf{J}$ is invertible. In that case, there exists a matrix left inverse $\mathbf{L}=(\mathbf{J}^H\mathbf{J})^{-1}\mathbf{J}^H$~\citep[see][chap.~4]{Strang_ILA2016}. Using the left inverse, the result is an estimate
\begin{eqnarray}
\tilde{\mathbf{e}}\tilde{\mathbf{e}}^H&=&\left(\mathbf{J}_1^H\mathbf{J}_1\right)^{-1}\mathbf{J}_1^H\mathbf{v}_1\mathbf{v}_2^H\mathbf{J}_2\left(\mathbf{J}^H_2\mathbf{J}_2\right)^{-H}\nonumber \\
&=&\mathbf{L}_1\mathbf{v}_1\mathbf{v}_2^H\mathbf{L}_2^H.
\label{eqn:pol}
\end{eqnarray}
SEFD is found by equating the flux density  to the standard deviation of the sum of the diagonal of the noisy estimate $\tilde{\mathbf{e}}\tilde{\mathbf{e}}^H$~(see Sec.~3 in S2021),
\begin{eqnarray}
\text{SEFD}&=&\frac{\text{SDev}\left[(\tilde{\mathbf{e}}\tilde{\mathbf{e}}^H)_{1,1}+(\tilde{\mathbf{e}}\tilde{\mathbf{e}}^H)_{2,2}\right]}{\eta_0}=\frac{\text{SDev}(\tilde{I})}{\eta_0},
\label{eqn:SEFD_STDev}
\end{eqnarray}
where $._{1,1}, ._{2,2}$ denote the diagonal entries of Eq.~(\ref{eqn:pol}) and $\eta_0\approx 120\pi\,\Omega$ is the free space impedance. We note that the approach discussed in this section does not involve specializations. It is valid for any system from dual-polarized to multipole antennas. The only requirement is that the columns of the Jones matrices are independent, which poses no practical complication. 

Therefore, the derivation of the SEFD expression entails working out Eq.~\eqref{eqn:SEFD_STDev} into a formula that involves the system temperatures, $T_{\rm sys}$, seen by each antenna and the physical properties of the antennas such as the effective lengths and antenna resistances. 
We summarize the steps as follows. First, we \added{recognized that the random variable $\tilde{I}$ in Eq.~\eqref{eqn:SEFD_STDev} is the trace of matrix $\tilde{\mathbf{e}}\tilde{\mathbf{e}}^H$}
\begin{eqnarray}
\tilde{I}=\text{tr}\left(\tilde{\mathbf{e}}\tilde{\mathbf{e}}^H\right)
=\text{tr}\left(\mathbf{L}_1\mathbf{v}_1\mathbf{v}_2^H\mathbf{L}_2^H\right).
\label{eqn:I_tilde}
\end{eqnarray}
\added{Applying the trace property of matrix product}\footnote{This can be shown by exchanging the summation order, for example see planetmath.org/proofofpropertiesoftraceofamatrix, www.statlect.com/matrix-algebra/trace-of-a-matrix, en.wikipedia.org/wiki/Trace\_(linear\_algebra)\#cite\_note-4}\added{, $\text{tr}(AB)=\text{tr}(BA)$ which is valid for $A~(m\times n)$ and $B~(n\times m$), to Eq.~\eqref{eqn:I_tilde}, leads to} 
\begin{eqnarray}
\tilde{I}
&=&\text{tr}\left(\mathbf{v}_1\mathbf{v}_2^H\mathbf{L}_2^H\mathbf{L}_1\right) =\text{tr}\left(\mathbf{v}_2^H\mathbf{L}_2^H\mathbf{L}_1\mathbf{v}_1\right) \nonumber \\
&=&\mathbf{v}_2^H\mathbf{L}_2^H\mathbf{L}_1\mathbf{v}_1.
\label{eqn:I_tilde2}
\end{eqnarray}
\added{The last equality  in Eq.~\eqref{eqn:I_tilde2} comes from the fact that $\mathbf{v}_2^H\mathbf{L}_2^H\mathbf{L}_1\mathbf{v}_1$ produces a single number such that the trace operator is redundant. Therefore, we can write Eq.~\eqref{eqn:I_tilde2} as}
\begin{eqnarray}
\tilde{I}
=\left[ \begin{array}{c c c }V_{X2}^* & V_{Y2}^* & V_{Z2}^*\end{array} \right]\mathbf{M}
\left[ \begin{array}{c}
V_{X1} \\
V_{Y1} \\
V_{Z1}
\end{array} \right],
\label{eqn:I_tilde3}
\end{eqnarray}
\added{where we introduce $\mathbf{M}=\mathbf{L}_2^H\mathbf{L}_1$ for brevity. Eq.~\eqref{eqn:I_tilde3} is convenient because $\mathbf{M}$ contains only antenna effective lengths that are deterministic quantities; the flanking vectors $\mathbf{v}_1,~\mathbf{v}_2$ represent the random variables of the noisy voltages.}

\added{Next, we compute the variance,   $\text{Var}(\tilde{I})$, of Eq.~\eqref{eqn:I_tilde3}}. This results in a lengthy sum that contains two components: the variance of the correlation of the antenna voltages \added{(for example, $\text{Var}(V_{X1}V_{X2}^*), \text{Var}(V_{X1}V_{Y2}^*)$, et cetera) with scalar coefficients} and the covariance of the cross terms with \added{different} scalar coefficients. \added{The scalar coefficients associated with the variances} come from the magnitude squared of the \added{entries of $\mathbf{M}$}. \added{The example in Sec.~\ref{sec:SEFD} illustrates the algebraic expansion of Eq.~\eqref{eqn:I_tilde3} and the appendix shows} that the variance terms contain the products of the antenna $T_{\rm sys}$'s and antenna resistances \added{(which represent the noisy voltages in Eq.~\eqref{eqn:I_tilde3})} and that the covariance of the cross terms vanishes. Finally, the standard deviation is the square root of the variance.

\section{General SEFD expression}
\label{sec:Gen_SEFD}
Following the steps described in the last paragraph in Sec.~\ref{sec:back}, it can be shown that the direction-dependent SEFD of a two-element tripole interferometer is found as
\begin{eqnarray}
\text{SEFD} = \frac{4k R_{\rm ant}}{ \eta_0}\sqrt{\mathbf{t}_{\rm sys1}^T(\mathbf{M}\circ \mathbf{M}^*)\mathbf{t}_{\rm sys2}}, 
\label{eqn:SEFD_gen}
\end{eqnarray}
with units \si{\watt\per\metre\squared \per \hertz}, where $\circ$ denotes the Hadamard (element-by-element) matrix product 
\added{such that $\mathbf{M}\circ \mathbf{M}^*$ produces the magnitude squared of each entry  of matrix $\mathbf{M}$}; $k$ is the Boltzmann constant; $\mathbf{t}_{\rm sys1}=[T_{{\rm sys}X1}, T_{{\rm sys}Y1}, T_{{\rm sys}Z1}]^T$ in \si{\kelvin} and similarly with $\mathbf{t}_{\rm sys2}$. $R_{\rm ant}$ in \si{\ohm} is the antenna resistance which we assume to be identical. Non-identical antenna resistances can be accounted for by a simple modification. 
\begin{eqnarray}
\text{SEFD} = \frac{4k}{ \eta_0}\sqrt{\mathbf{t}_{\rm sysR1}^T(\mathbf{M}\circ \mathbf{M}^*)\mathbf{t}_{\rm sysR2}}, 
\label{eqn:SEFD_gen2}
\end{eqnarray}
where
\begin{eqnarray}
\mathbf{t}_{\rm sysR1}=\left[ \begin{array}{c}
T_{{\rm sys}X1}R_{{\rm ant}, X1}\\
T_{{\rm sys}Y1}R_{{\rm ant}, Y1} \\
T_{{\rm sys}Z1}R_{{\rm ant}, Z1}
\end{array} \right],
\label{eqn:nonID_t}
\end{eqnarray}
and similarly for $\mathbf{t}_{\rm sysR2}$ (with entries that pertain to $T_{sys}$'s and $R_{ant}$'s of system 2). 

Although Eq.~\eqref{eqn:SEFD_gen} may look unfamiliar, we can make connections to previously known results. First, for dual polarized antennas, it can be shown  that Eq.~\eqref{eqn:SEFD_gen} produces identical SEFD to~Eq.~(20) in~S2021. This may be accomplished by replacing the Jones matrix in Eq.~\eqref{eqn:outer_voltage} with a $2\times2$ matrix representing the effective lengths of the dual-polarized antennas and following through the steps. In this case, the left inverse is the matrix inverse. Second, in the special case of diagonal matrix $\mathbf{M}$, the SEFD only contains square terms such as $T_{{\rm sys}X}^2, T_{{\rm sys}Y}^2$, etc. which leads to the approximation in Eq.~\eqref{eqn:SEFD_XYapprox}. In general, however, we obtain cross terms such as $T_{{\rm sys}X}T_{{\rm sys}Y}$, etc. 

The term under the square root sign in Eq.~\eqref{eqn:SEFD_gen} can be written as~\citep{Million_paper}
\begin{eqnarray}
\mathbf{t}_{\rm sys1}^T(\mathbf{M}\circ\mathbf{M}^*)\mathbf{t}_{\rm sys2}=\mathrm{tr}\left(\mathbf{D}_{t1}\mathbf{M}\mathbf{D}_{t2}\mathbf{M}^H\right),
\label{eqn:tSSt_gen}
\end{eqnarray}
where $\mathbf{D}_{t1},\mathbf{D}_{t2}$ are diagonal matrices with elements of vectors $\mathbf{t}_{\rm sys1},\mathbf{t}_{\rm sys2}$ in the diagonals, respectively. This expression will become useful in estimating the behaviour of the quantity under the square root sign. \added{In Sec.~\ref{sec:SEFD}}, we consider a few special cases to illustrate the formula, its derivation and infer key insights.


\section{SEFD for a short tripole antenna system}
\label{sec:SEFD}
We consider a special case for short tripole antennas which are typical for an observation at very long wavelengths~\citep{Chen_2018ExA....45..231C, Chen_doi:10.1098/rsta.2019.0566}. Although the resulting expression is specialized to this case, we note that the process of derivation has much in common with the general case and helped us identify the general formula. Let the tripole be of an identical effective length $\Delta l$ and coincide with the $x,y,z-$axes, respectively. In this case, the Jones matrix is given by
\begin{eqnarray}
\mathbf{J}_{s}&=&\Delta l \mathbf{Q}_{s}
=\Delta l \left[ \begin{array}{c c}
\mathbf{q}_{\theta s} & \mathbf{q}_{\phi s}\end{array} \right], \nonumber \\
&=&
\Delta l\left[ \begin{array}{cc}
\cos\theta \cos\phi & -\sin\phi\\
\cos\theta \sin\phi & \cos \phi \\
-\sin\theta & 0 \\
\end{array} \right],
\label{eqn:J_tri}
\end{eqnarray} 
where $\theta, \phi$ are the angles defined in the standard spherical coordinate system~(see S2021). The key attribute independent of the coordinate system is that the columns of $\mathbf{Q}_{s}$ are orthonormal: $\mathbf{q}_{\theta s}^T\mathbf{q}_{\phi s}=0, \norm{\mathbf{q}_{\theta s}}=\norm{\mathbf{q}_{\phi s}}=1$. Also, $\mathbf{Q}_s^T\mathbf{Q}_s=\mathbf{I}$, that is $\mathbf{Q}_s^T$ is the left inverse of $\mathbf{Q}_s$. The left inverse of $\mathbf{J}_s$ is
\begin{eqnarray}
\mathbf{L}_s=(\mathbf{J}_s^T\mathbf{J}_s)^{-1}\mathbf{J}_s=\frac{1}{\Delta l}\mathbf{Q}_s^T.
\label{eqn:Ls}
\end{eqnarray}

We consider a simple arrangement  where $X_1,X_2$ in the two antenna systems are parallel, and similarly with $Y_1, Y_2$ and $Z_1, Z_2$. In this case,
\begin{eqnarray}
\mathbf{M}_s=\mathbf{L}_s^T\mathbf{L}_s=\frac{1}{\Delta l^2}\mathbf{Q}_s\mathbf{Q}_s^T
\label{eqn:M_short}
\end{eqnarray}
is a symmetric real matrix; we note also that $\mathbf{Q}_s\mathbf{Q}_s^T=\mathbf{P}_s$ is a projection matrix~\citep[see][chap.~4]{Strang_ILA2016}.
The resulting variance (suppressing the leading $V_{\_}$ for brevity) is,
\begin{eqnarray}
\Delta l^4\text{Var}(\tilde{I}) &=& a^2\text{Var}(X_1X_2^*)+b^2\text{Var}(Y_1Y_2^*)\nonumber \\
&+&c^2\text{Var}(Z_1Z_2^*) \nonumber\\
&+&\left[\text{Var}(X_1Y_2^*) +\text{Var}(Y_1X_2^*)\right]d^2 \nonumber \\ &+&\left[\text{Var}(X_1Z_2^*)+\text{Var}(Z_1X_2^*)\right]e^2 \nonumber\\
&+&\left[ \text{Var}(Z_1Y_2^*)+\text{Var}(Y_1Z_2^*) \right]f^2
+2C,
\label{eqn:varI}
\end{eqnarray}
where $a=\mathbf{P}_{s1,1}, b=\mathbf{P}_{s2,2}, c=\mathbf{P}_{s3,3}; d=\mathbf{P}_{s1,2}; e=\mathbf{P}_{s1,3}; f=\mathbf{P}_{s2,3}$ are the entries of matrix $\mathbf{P}_{s}$; $C$ is the covariance of cross terms. More detailed discussion regarding the right hand side of Eq.~\eqref{eqn:varI} is given in the appendix. The important conclusions are that $C=0$ and only the $\mathrm{Var}()$ terms remain. We further note that these $\mathrm{Var}()$ terms equate to terms that contain a mixture of system temperatures and antenna resistances, as shown in Eq.~\eqref{eqn:vars}. This result is achieved only with two fundamental assumptions: the mutual coherence seen by antenna systems 1 and 2 due to the system noise is negligible and the noise is zero mean and has iid real and imaginary components. 

The resulting SEFD for short orthogonal tripoles is
\begin{eqnarray}
\text{SEFD}_{s} &=& \frac{4k R_{\rm ant}}{\Delta l^2 \eta_0}\sqrt{\mathbf{t}_{\rm sys}^T(\mathbf{P}_s\circ\mathbf{P}_s)\mathbf{t}_{\rm sys}}, \nonumber \\
&=& \frac{8\pi k}{3\lambda^2}\sqrt{\mathbf{t}_{\rm sys}^T(\mathbf{P}_s\circ\mathbf{P}_s)\mathbf{t}_{\rm sys}},
\label{eqn:SEFDs}
\end{eqnarray}
where we assume antennas of an identical design and the same sky illumination for system 1 and 2. However, the $X,Y,Z$ antennas place different emphases on the sky which may result in differing $T_{sys}$'s. The second line comes from $R_{\rm ant}=80\pi^2(\Delta l/\lambda)^2$ for a short dipole where $\Delta l$ is the effective length~\citep[see][chap.~11]{Cheng_FW1992}. Therefore, 
\begin{eqnarray}
\frac{R_{\rm ant}}{\Delta l^2\eta_0}=\frac{2\pi}{3\lambda^2}. 
\label{eqn:Rant_delt2}
\end{eqnarray}

We readily see the connection between the top line of Eq.~\eqref{eqn:SEFDs} and the  Eq.~\eqref{eqn:SEFD_gen}. The conjugate sign in the second $\mathbf{M}^*$ in Eq.~\eqref{eqn:SEFD_gen} is needed since $\mathbf{M}$ is generally a complex matrix, but in the case of the short dipoles, $\mathbf{M}$ becomes real.

\subsection{Special case: Identical $T_{\rm sys}$}
\label{sec:id_Tsys}
Next, we consider a special case where the system temperatures
of the $X,Y,Z$ antennas are equal, $T_{{\rm sys}X}=T_{{\rm sys}Y}=T_{{\rm sys}Z}$, such that $\mathbf{t}_{\rm sys}=T_{\rm sys}[1,1,1]^T$. This leads to
\begin{eqnarray}
\text{SEFD}_{s:id-T_{\rm sys}} &=& \frac{8\pi k T_{\rm sys}}{3\lambda^2}\sqrt{\text{tr}(\mathbf{P}_s^2)},\nonumber\\
&=& \frac{8\pi k T_{\rm sys}}{3\lambda^2}\sqrt{2},
\label{eqn:SEFDs_syseq}
\end{eqnarray}
The first line comes from application of Eq.~\eqref{eqn:tSSt_gen}. The second line comes from the following reasoning. A property of projection matrix is  $\mathbf{P}_s^2=\mathbf{P}_s$~\citep[see][chap.~4]{Strang_ILA2016}. Therefore $\text{tr}(\mathbf{P}_s^2)=\text{tr}(\mathbf{P}_s)$, where $\text{tr}()$ indicates the trace of the matrix, which is the sum of the main diagonal entries. It is also known that trace of a matrix is the sum of its eigenvalues~\citep[see][chap.~6]{Strang_ILA2016}. Writing
\begin{eqnarray}\mathbf{P}_s&=&\mathbf{Q}_s\mathbf{Q}_s^T=1\mathbf{q}_{\theta s}\mathbf{q}_{\theta s}^T+1\mathbf{q}_{\phi s}\mathbf{q}_{\phi s}^T+0\mathbf{q}_{\perp s}\mathbf{q}_{\perp s}^T\nonumber \\
&=&\mathbf{Q}\Lambda \mathbf{Q}^T,
\label{eqn:spectral}
\end{eqnarray}
suggests it has eigenvalues in $\Lambda=\mathrm{diag}(1, 1, 0)$, such that the sum is 2; $\mathbf{q}_{\perp s}$ is an orthonormal eigenvector perpendicular to $\mathbf{q}_{\theta s}$ and $\mathbf{q}_{\phi s}$; $\mathbf{Q}=\left[\mathbf{q}_{\theta s}~ \mathbf{q}_{\phi s}~\mathbf{q}_{\perp s} \right]$ is an orthogonal matrix.

We note that $\text{SEFD}_{s:id-T_{\rm sys}}$ in equation~\eqref{eqn:SEFDs_syseq} is direction-independent. This is expected because, unlike dual-polarized dipoles or a single dipole, the tripole system has no blind spot. In fact, because the columns of $\mathbf{Q}_s$ are orthonormal, the tripole system does not scale the incident electric field in a way that is direction dependent. We can see this by finding the length of voltage vector $\mathbf{v}=\mathbf{J}_s\mathbf{e}=\Delta l\, \mathbf{Q}_s\mathbf{e}$,
\begin{eqnarray}
\mathbf{v}^H\mathbf{v}=\norm{\mathbf{v}}^2=\Delta l^2\mathbf{e}^H\mathbf{Q}_s^T\mathbf{Q}_s\mathbf{e}=\Delta l^2 \norm{\mathbf{e}}^2.
\label{eqn:norm_v}
\end{eqnarray}
since $\mathbf{Q}_s^T\mathbf{Q}_s=\mathbf{I}$. This finding is fully consistent with the approach using singular values~\citep{Carozzi_7297193}. 

\subsection{Comparison to dual-polarized dipoles and single dipole}
\label{sec:comp2_1}
It is instructive to compare Eq.~\eqref{eqn:SEFDs_syseq} with $X,Y$ dual-polarized dipoles and a single dipole. For the $X,Y$ dipoles, we can show that~(see S2021)
\begin{eqnarray}
\text{SEFD}_{sXY:id-T_{\rm sys}} &=& \frac{8\pi k T_{\rm sys}}{3\lambda^2}\sqrt{\frac{1}{\cos^4\theta}+1},
\label{eqn:SEFDsXY_syseq}
\end{eqnarray}
where $\theta$ is the angle with respect to the $z$-axis. Hence, in all directions, the tripole system achieves the minimum SEFD of the $X,Y$ system. For the single $Z$ dipole, the antenna effective area is~\citep[see][chap.~11]{Cheng_FW1992}
\begin{eqnarray}
A_e=\frac{3\sin^2\theta}{2}\frac{\lambda^2}{4\pi}\label{eqn:Aez}
\end{eqnarray}
Assuming an unpolarized source, 
\begin{eqnarray}
\text{SEFD}_{sZ}=k\frac{T_{\rm sys}}{A_e/2} = \frac{8\pi k T_{\rm sys}}{3\lambda^2}\frac{2}{\sin^2\theta},
\label{eqn:SEFD_Z}
\end{eqnarray}
which is at best $\sqrt{2}$ times the lowest value of Eq.~\eqref{eqn:SEFDsXY_syseq} and Eq.~\eqref{eqn:SEFDs_syseq}. Equations~\eqref{eqn:SEFDs_syseq}, ~\eqref{eqn:SEFDsXY_syseq}, and ~\eqref{eqn:SEFD_Z} provide a quantitative basis for comparing the performance of tripole systems to dual-polarized and single antenna systems assuming identical $T_{\rm sys}$. These expressions are comparable to estimates found in, for example ~\citet{ZARKA2012156}, but are based on more detailed reasoning.

\section{Example: Tripole antenna for long wavelength observation}
\label{sec:Example}
\begin{figure}[htb]
\begin{center}
\noindent
  \includegraphics[width=3in]{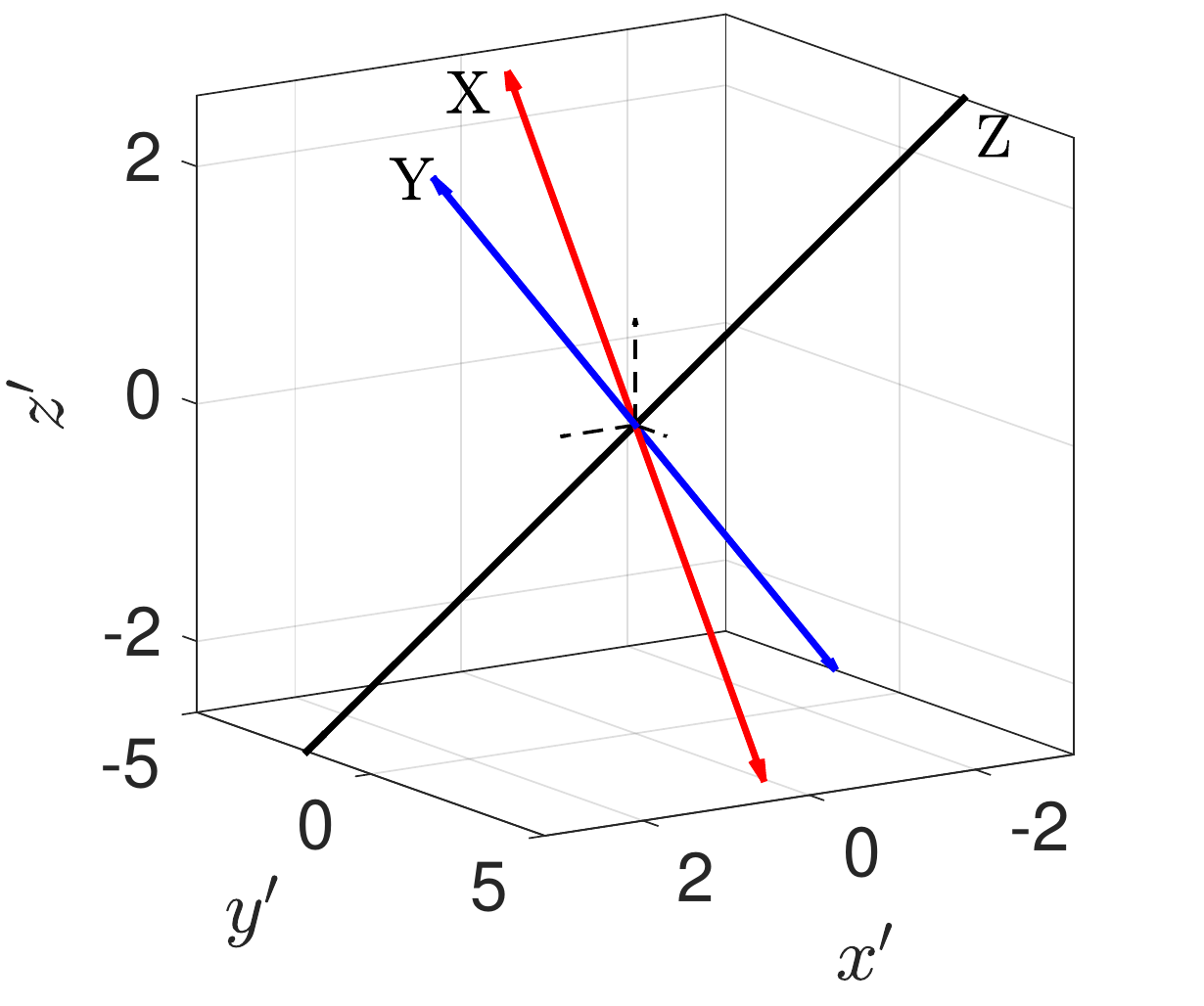}
\caption{Orthogonal tripole system. Each dipole is  \SI{5}{\metre} long. The dashed lines at the origin are unit vectors in the $+x',+y',+z'$ directions. The coordinates of the tips of the dipoles in the $+z'$ half space are $2.5(-1/\sqrt{2},1/\sqrt{6},1/\sqrt{3})$ (black), $2.5(0,-2/\sqrt{6},1/\sqrt{3})$ (red), and $2.5(1/\sqrt{2},1/\sqrt{6},1/\sqrt{3})$ (blue), respectively.}
\label{fig:Tripole_dsl}
\end{center}
\end{figure}

In this section, we demonstrate SEFD expressions given in Eq.~\eqref{eqn:SEFD_gen} and Eq.~\eqref{eqn:SEFDs} applied to the tripole antenna shown in Fig.~\ref{fig:Tripole_dsl}. The tripole system is representative of that being considered for the DSL lunar orbiting interferometer~\citep{Chen_2018ExA....45..231C, Chen_doi:10.1098/rsta.2019.0566}. The dipoles are \SI{5}{\metre} long (half wavelength at \SI{30}{\mega\hertz}) and  are geometrically mutually orthogonal. Since the intended frequency range is \SI{0.1}{\mega\hertz} to \SI{30}{\mega\hertz} , we performed the analysis at \SI{3}{\mega\hertz}, \SI{10}{\mega\hertz} and \SI{30}{\mega\hertz}. The aim of our current work is to demonstrate the SEFD formulas, how to apply and use them, and establish that they are self-consistent. Calculation of the \added{imaging} sensitivity of the DSL mission\added{, which would involve the interferometer as a whole and the orbital parameters including the effects of the Moon \citep{ShiImaging2021}, } is well beyond our current scope.

We computed the SEFD using the general expression with Eq.~\eqref{eqn:SEFD_gen} and the analytical formula for assuming orthogonal short tripoles with Eq.~\eqref{eqn:SEFDs}.  For the general formula, we generate the antenna Jones matrix using full-wave electromagnetic simulation which makes no assumption regarding the orthogonality nor whether the antennas are short. In fact, more details regarding the body of the satellite were included in the model. We expect the results for the general expression and the short dipole to converge at frequencies well below \SI{30}{\mega\hertz} where the tripoles are indeed short relative to observation wavelengths. 

\begin{figure}[htb]
\begin{center}
\noindent
  \includegraphics[width=0.4\textwidth]{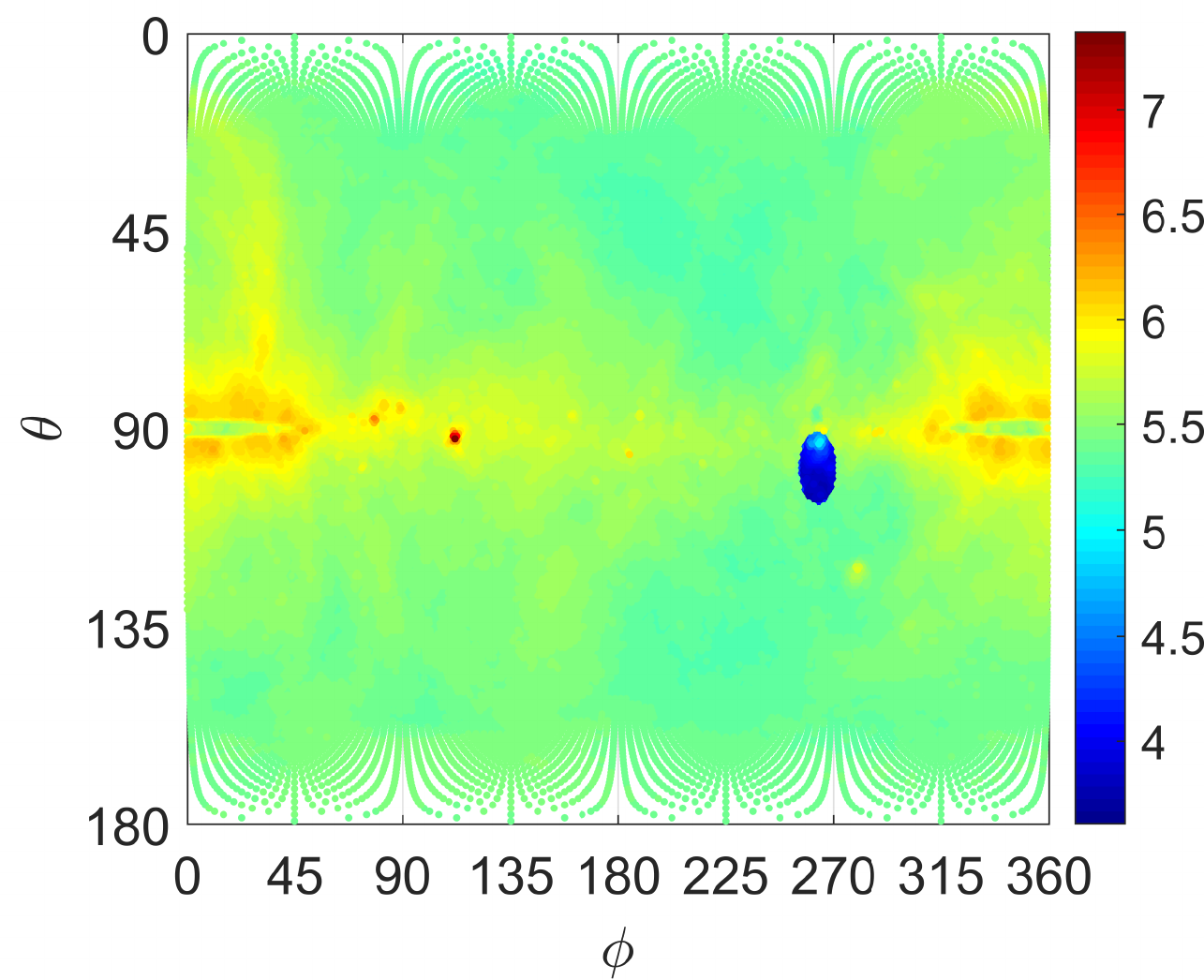}
\caption{$\log_{10}(T_{\rm sky})$ at \SI{10}{\mega\hertz} based on ULSA, displayed in Galactic Coordinates.}
\label{fig:10MHz_sky}
\end{center}
\end{figure}

For this example, we assumed $\mathbf{t}_{{\rm sys}1}=\mathbf{t}_{{\rm sys}2}$ but $T_{{\rm sys}X},T_{{\rm sys}Y}, T_{{\rm sys}Z}$ were different due to the anisotropic sky. This is consistent with a snapshot observation for baseline distances of approximately \SI{10}{\kilo\metre} or less (for orbital height of approximately \SI{300}{\kilo\metre} above the lunar surface).  We used the Ultralong-wavelength Sky Model with Absorption effect (ULSA\footnote{https://github.com/Yanping-Cong/ULSA}) described in ~\citet{Cong_2021ApJ...914..128C} at \SI{3}{\mega\hertz}, \SI{10}{\mega\hertz}, and \SI{30}{\mega\hertz} with constant spectral index model with enhanced fluctuation parameter, $F_1 = 3$ \citep[see][Fig.~11]{Cong_2021ApJ...914..128C}. As an example, the sky map at \SI{10}{\mega\hertz} is shown in Fig.~\ref{fig:10MHz_sky}.

\subsection{Orthogonal short tripole model}
\label{sec:orto_short}
\begin{figure}[t!]
\begin{center}
\noindent
  \includegraphics[width=0.45\textwidth]{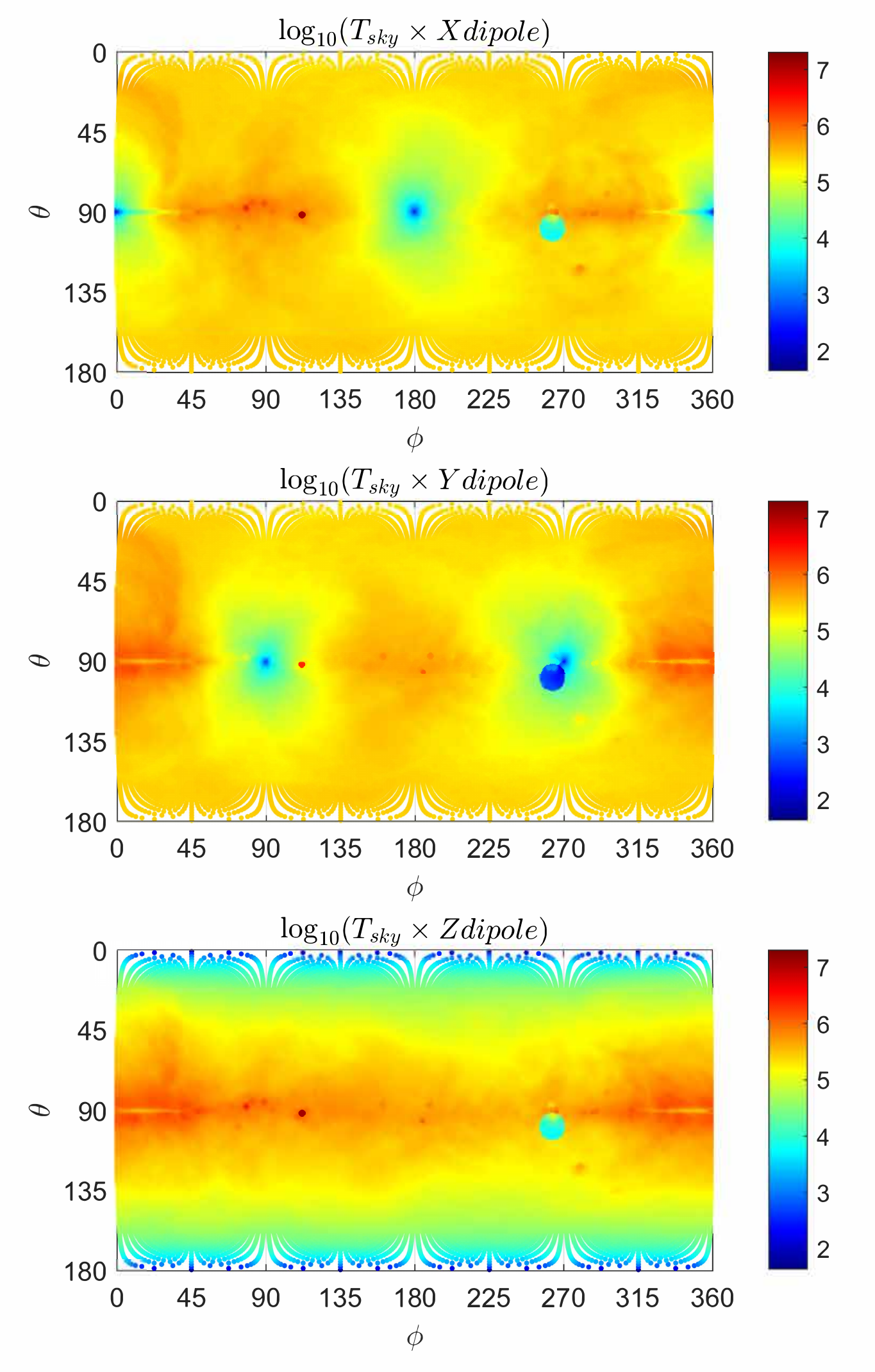}
\caption{Sky temperature at \SI{10}{\mega\hertz} as seen by the normalized power pattern of the dipoles, $\log_{10}(T_{\rm sky}\times\text{[X, Y, Z]~dipole})$, displayed in Galactic Coordinates.}
\label{fig:XYZ_10MHz_sky}
\end{center}
\end{figure}

In this example, the $X, Y, Z$ dipoles are oriented such that $+X$ dipole points to the Galactic center at $\theta=90\degree,\phi=0\degree$ in Fig.~\ref{fig:10MHz_sky}. Consequently, $+Y$ dipole points to $\theta=90\degree,\phi=90\degree$, and $+Z$ dipole points to $\theta=0\degree$.
Here, we assume a fixed orientation of the tripoles relative to the sky, which may be thought of as a snapshot tripole sensitivity for illustrative purposes. Note that for the DSL mission, the satellites are tentatively oriented with respect to the Moon center, so that  the changing pointing of the tripoles when orbiting the Moon results in a more balanced sensitivity for different directions in the sky. 
Continuing with our example, each dipole sees a different sky as shown in Fig.~\ref{fig:XYZ_10MHz_sky} which leads to differing antenna temperatures. This was obtained by integrating the normalized power pattern of the antennas with the sky temperature distribution~\citep[see][chap.~17]{Kraus}. The resulting antenna temperatures at \SI{10}{\mega\hertz} are 
$T_{{\rm ant}X} = \SI{323}{\kilo\kelvin}, T_{{\rm ant}Y}=\SI{359}{\kilo\kelvin}, T_{{\rm ant}Z}=\SI{400}{\kilo\kelvin}$.

Based on the preliminary design of the DSL low frequency interferometer system, we assume a receiver (rx) root mean square (rms) voltage noise of $V_{\rm rx:rms}=3\sqrt{2}\,\text{nV\,Hz}^{-0.5}$. \added{This $V_{\rm rx:rms}$ is the equivalent open-circuit noise voltage present at the antenna port due to the receiver noise.}
The corresponding receiver noise temperature, $T_{\rm rx}$, may be computed based on $V_{\rm rx:rms}=\sqrt{4kT_{\rm rx}R_{\rm ant}}$~\citep[see][append.~K]{Gonzalez_1997}. For $R_{\rm ant}$, we assume a triangular current distribution such that the effective length is one half that of the physical length, $\Delta l=\SI{2.5}{\metre}$. This results in $T_{\rm rx}=\SI{59.4}{\kilo\kelvin}$ at \SI{10}{\mega\hertz}. The system temperature was taken as the sum, $T_{\rm ant}+T_{\rm rx}$, for each antenna.

\begin{figure}[htb]
\begin{center}
\noindent
\includegraphics[width=0.4\textwidth]{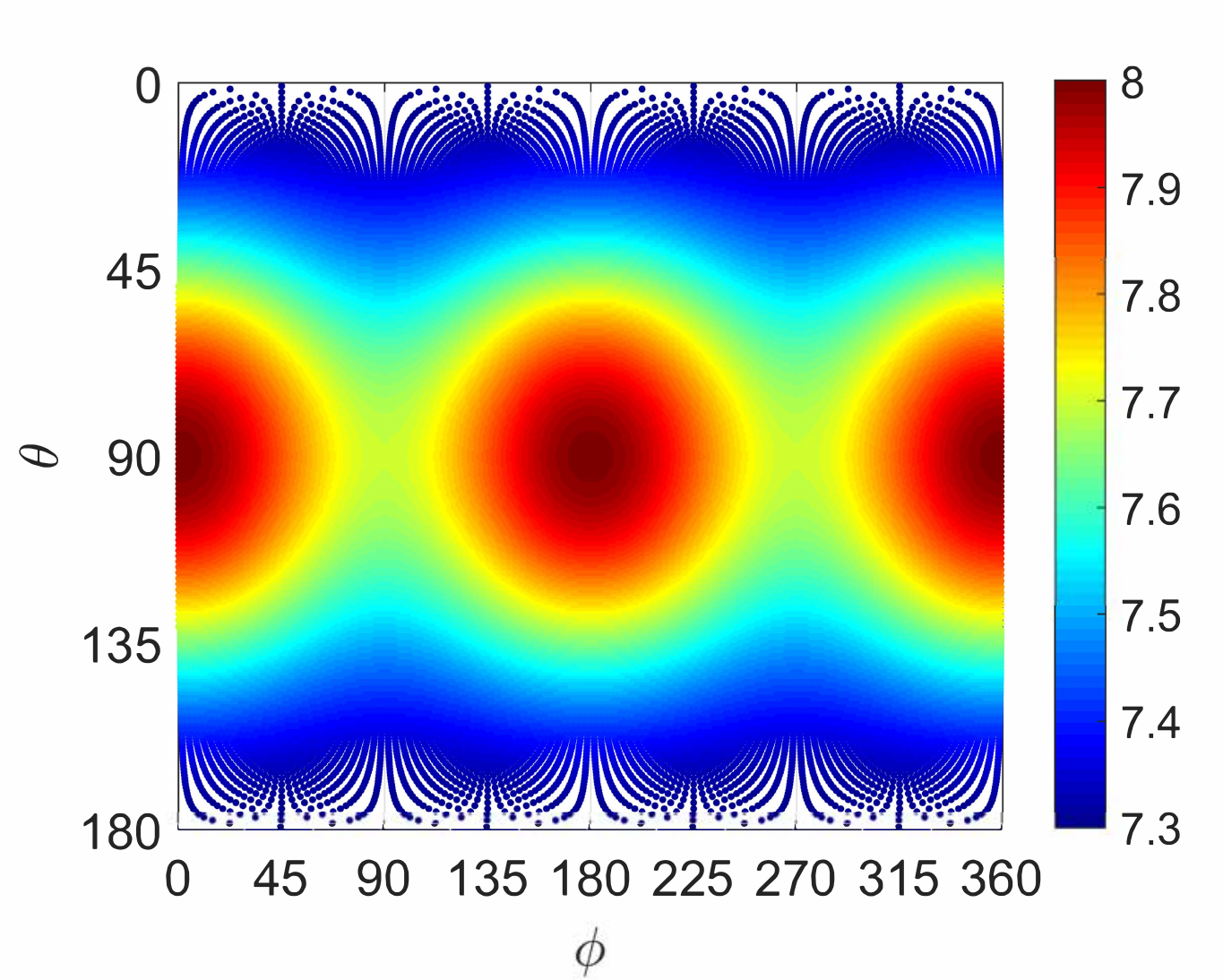}
\caption{SEFD$(\theta,\phi)$ in MJy at \SI{10}{\mega\hertz} for the orthogonal tripole, computed using Eq.~\eqref{eqn:SEFDs}.}
\label{fig:SEFD_10MHz_ae}
\end{center}
\end{figure}

The SEFD was computed using Eq.~\eqref{eqn:SEFDs} and is shown in Fig.~\ref{fig:SEFD_10MHz_ae}. At \SI{10}{\mega\hertz}, the SEFD varies from 7.3\,MJy to 8\,MJy. This slight variation is the result of the unequal $T_{sysX}, T_{sysY}, T_{sysZ}$ interacting with the $\mathbf{P}_s\circ \mathbf{P}_s$ matrix. Fig.~\ref{fig:SEFD_10MHz_ae} suggests that for this orientation, the tripole system is less sensitive on the XY-plane and more sensitive in the direction of the Z-axis $(\phi=0\degree, 180\degree)$. This coincides with the Z-dipole being illuminated by the Galactic plane resulting in $T_{sysZ}$ being the highest. We repeated this calculation at \SI{3}{\mega\hertz} and  \SI{30}{\mega\hertz}. At  \SI{3}{\mega\hertz}, 
$T_{{\rm ant}X} = \SI{5283}{\kilo\kelvin}, T_{{\rm ant}Y}=\SI{5452}{\kilo\kelvin}, T_{{\rm ant}Z}=\SI{ 5463}{\kilo\kelvin}$, which are nearly equal, and $T_{\rm rx}=\SI{660}{\kilo\kelvin}$. The SEFD has a very similar pattern to Fig.~\ref{fig:SEFD_10MHz_ae}, however, the SEFD varies only very slightly, from approximately 9.9\,MJy to 10\,MJy. This is expected because the $T_{\rm sys}$ are nearly equal. At  \SI{30}{\mega\hertz}, 
$T_{{\rm ant}X} = \SI{20.6}{\kilo\kelvin}, T_{{\rm ant}Y}=\SI{23.9}{\kilo\kelvin}, T_{{\rm ant}Z}=\SI{ 27.1}{\kilo\kelvin}$, and $T_{\rm rx}=\SI{6.6}{\kilo\kelvin}$. The resulting SEFD varies from approximately 4.7\,MJy to 5.3\,MJy following a very similar pattern as Fig.~\ref{fig:SEFD_10MHz_ae}, and thus, are not shown here.

\begin{table*}[th]
\centering
\caption{Antenna temperature and SEFD for 3\,MHz, 10\,MHz and 30\,MHz.} \label{tbl:compar}
\renewcommand{\arraystretch}{1.3}
\begin{tabular}{l l l l l l l m{1.7cm}}
\cline{2-8}
                             & \boldmath$R_{\rm ant}$ [\si{\ohm}] & \boldmath$T_{\rm rx}$ [kK] & \boldmath$T_{{\rm ant}X}$ [kK] & \boldmath$T_{{\rm ant}Y}$ [kK] & \boldmath$T_{{\rm ant}Z}$ [kK] & \textbf{SEFD} [MJy] & \boldmath$\Delta_{\rm SEFD}$, \% \\ \hline 
                             & \multicolumn{7}{c}{Analytical tripole antenna (short dipole approximation)} \\ \hline
\multicolumn{1}{ l }{3 MHz}  & 0.5                & 660            & 5283                 & 5452               & 5463                 & 9.9 to 10           & 1.5                     \\ 
\multicolumn{1}{ l }{10 MHz} & 5.5                & 59.4             & 324                & 359                & 400                  & 7.3 to 8.0            & 9.1                     \\ 
\multicolumn{1}{l}{30 MHz}  & 49.4               & 6.6              & 20.6                & 23.9              & 27.1                & 4.7 to 5.3        & 10.7                    \\ \hline
                             & \multicolumn{7}{c}{Simulated tripole antenna (5 m long)}                                                                                                   \\ 
                             \hline
\multicolumn{1}{l}{3 MHz}  & 0.49 -- $j$4442        & 668            & 5289                 & 5469               & 5478                 & 9.9 to 10           & 1.5                     \\ 
\multicolumn{1}{l}{10 MHz} & 5.7 -- $j$1209         & 57.5             & 327                & 363                & 404                  & 7.3 to 8.0          & 9.1                     \\ 
\multicolumn{1}{l}{30 MHz} & 81 -- $j$43            & 4.04             & 20.2                & 23.9                 & 28.0                   & 3.9 to 5.0         & 23.6                    \\ \hline
                             & \multicolumn{7}{c}{Simulated tripole antenna (5 m long) with the satellite (1.154\,m x 1.086\,m x 0.29\,m)}                                                       \\ 
                             \hline
\multicolumn{1}{l}{3 MHz}  & 0.37 -- $j$4160       & 881            & 529                 & 5459               & 5481                & 10.1 to 10.8         & 6.7                     \\ 
\multicolumn{1}{l}{10 MHz} & 4.4 -- $j$1140        & 74.1           & 324               & 359               & 400                 & 7.6 to 8.4          & 10.6                    \\ 
\multicolumn{1}{l}{30 MHz}  & 78 + $j$28           & 4.2             & 20.2               & 23.9               & 27.9                 & 4.0 to 5.2          & 24.7                    \\ \hline
\end{tabular}
\end{table*}

For comparison, we consider approximations based on Eq.~\eqref{eqn:SEFDs_syseq} assuming identical $T_{\rm sys}=T_{\rm avg}+T_{\rm rx}$, where $T_{\rm avg}$ is the average sky temperature based on the HEALPix pixel values of the sky map. At \SI{3}{\mega\hertz}, $T_{\rm avg}=\SI{5400}{\kilo\kelvin}$ resulting in $\text{SEFD}_{s:id-T_{\rm sys}}=9.92$\,MJy, which is very similar to our result because the $T_{\rm sys}$ seen by $X, Y, Z$ dipoles are nearly equal. At \SI{10}{\mega\hertz}, $T_{\rm avg}=\SI{361}{\kilo\kelvin}$ and $\text{SEFD}_{s:id-T_{\rm sys}}=\num{7.65}$\,MJy, which is the midpoint between our result of 7.3\,MJy to 8\,MJy. This is similarly the case at \SI{30}{\mega\hertz}, where $T_{\rm avg}=\SI{   23.9}{\kilo\kelvin}$ and $\text{SEFD}_{s:id-T_{\rm sys}}=\num{4.99}$\,MJy, which is also the average of 4.7\,MJy and 5.3\,MJy.



\subsection{Tripole model with electromagnetic simulation}
\label{sec:EM_tripole}

\begin{figure}[b!]
\begin{center}
\noindent
  \includegraphics[width=3in]{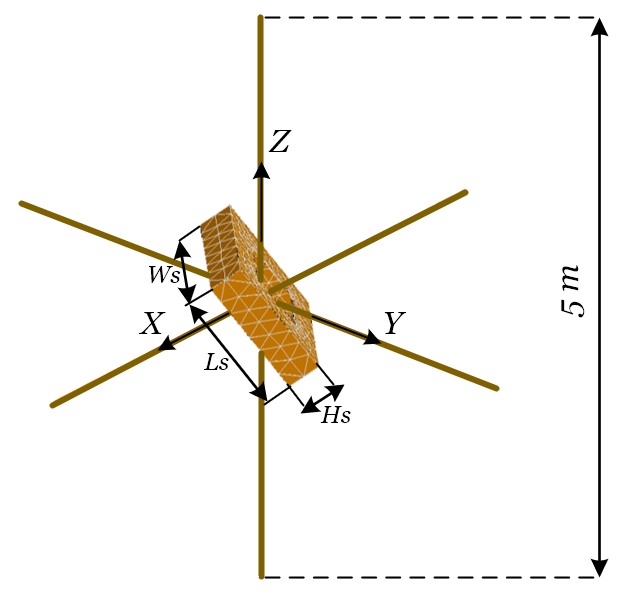}
\caption{Tripole antenna with the daughter satellite dimensions $(L_{s}, W_{s}, H_{s})=(1.154, 1.086, 0.29)$~m simulated in FEKO Altair.}
\label{fig:tripole_w_sat}
\end{center}
\end{figure}

In this section, we provide the results of SEFD$(\theta, \phi)$ obtained from Eq.~\eqref{eqn:SEFD_gen} using simulated results from the computational electromagnetic software package FEKO Altair. We compare the simulated SEFD range and its percentage spread with the analytical results in Table~\ref{tbl:compar}. First, the tripole was simulated on its own, and then, with the presence of metallic body of the daughter satellite, as shown in Fig.~\ref{fig:tripole_w_sat}. These simulations help us distinguish the contribution of the satellite body to any measurable difference detected in the results from that due to the difference between the  analytical expression and the electromagnetic simulation alone. The dipoles in each polarization were 5~m long, and the satellite with dimensions $(L_{s}, W_{s}, H_{s})=(1.154, 1.086, 0.29)$~m was tilted by the angle $\alpha = \arccos(1/\sqrt(3)) \approx 54.7\degree $ off $z$-axis so that $X-$dipole aligned with the Galaxy centre. We used simulated input impedance and normalized power patterns to define the effective lengths that construct the Jones matrix~\citep{Ung2020}. At each frequency, the effective lengths of the tripole were calculated as
\begin{eqnarray}
\mathbf{J} & = & -j\frac{2 \lambda }{\eta_0} Z_{ant}
\left[ \begin{array}{cc}
E_{X\theta} & E_{X\phi} \\
E_{Y\theta} & E_{Y\phi} \\
E_{Z\theta} & E_{Z\phi} \\
\end{array} \right],
\label{eqn:eff_len}
\end{eqnarray} 
where $\lambda$ is the wavelength of the frequency of interest, $\eta_{0}$ is the free space impedance, $Z_{ant}$ is the complex input impedance of the tripole and $E_{\theta}$, $E_{\phi}$ are the far-field components of the electric field. Electromagnetic simulation in FEKO was set up to separately excite X, Y and Z dipole antennas with 1~V source.  

\begin{figure}[t!]
\begin{center}
\noindent
\includegraphics[width=0.48\textwidth]{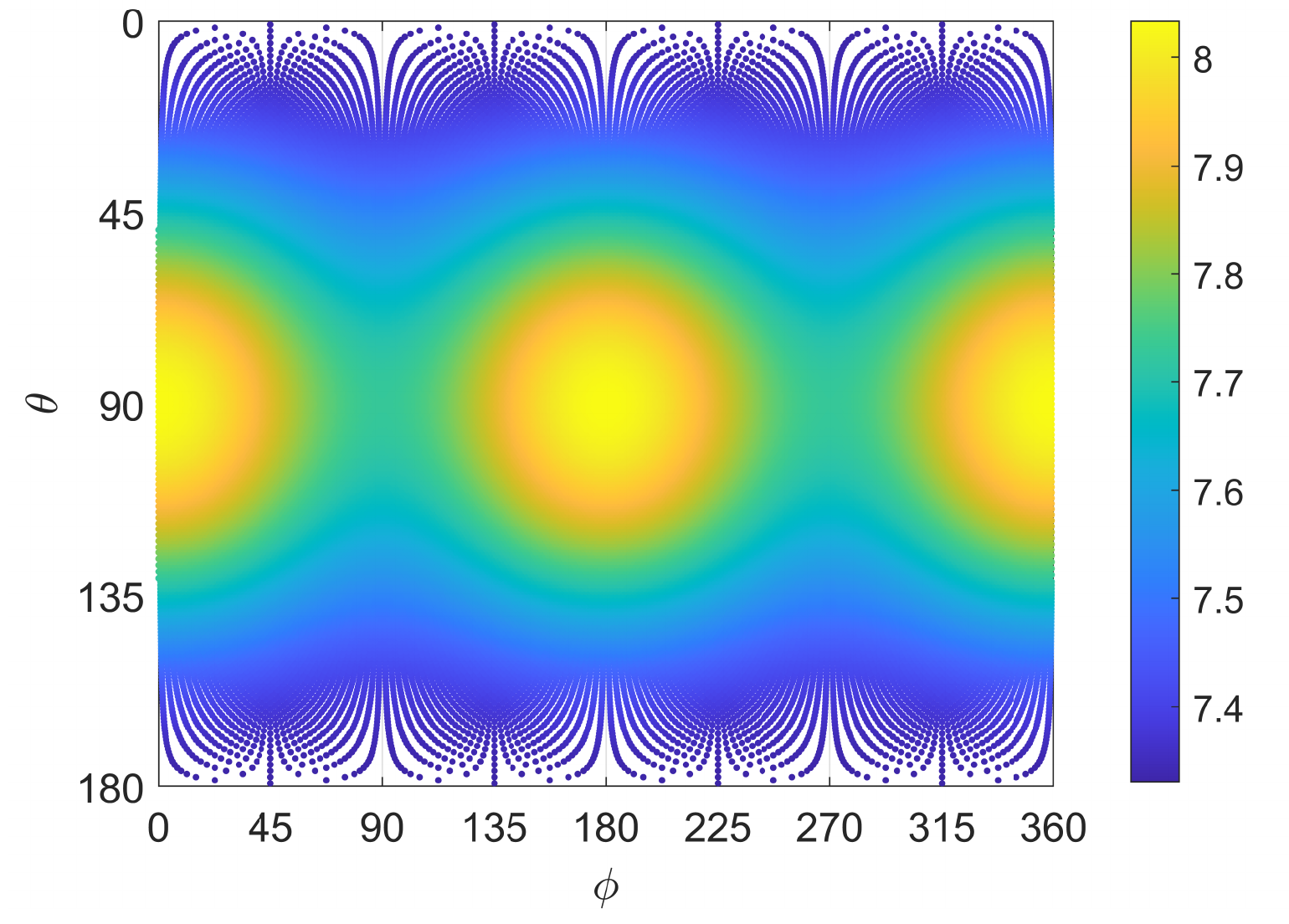}
\caption{SEFD$(\theta,\phi)$ in MJy at \SI{10}{\mega\hertz} for the orthogonal tripole without the satellite body, computed using Eq.~\eqref{eqn:SEFD_gen}.}
\label{fig:SEFD_10MHz_sim}
\end{center}
\end{figure}

Table~\ref{tbl:compar} summarizes the results. First, we compared the analytical result to the simulated tripole without the satellite body. We found that the analytical tripole represents the SEFD$(\theta, \phi)$ response very accurately at 3\,MHz and 10\,MHz, where the tripole is indeed a short dipole. Fig.~\ref{fig:SEFD_10MHz_sim} shows SEFD at 10\,MHz calculated using Eq.~\eqref{eqn:SEFD_gen}. As expected, SEFD values computed using simulated radiation patterns of a tripole at these frequencies have negligible discrepancy. We observe convergence of Eq.~\eqref{eqn:SEFDs} that is derived for analytical dipole and Eq.~\eqref{eqn:SEFD_gen} that is a general SEFD expression. At 30\,MHz, where the dipoles are half-wavelength long, the approximation is still reasonable, but there is a measurable difference in SEFD. Therefore, for antenna systems with unknown closed-form expressions, such as the half-wavelength dipole, Eq.~\eqref{eqn:SEFD_gen} combined with electromagnetic simulation results should be used for highest accuracy. The simulated tripole with the satellite body  produced some measurable differences in SEFD to both the simulated tripole without the body and analytical result, though the latter produced sufficiently accurate estimates. However, if a precise design study is required, the presence of the satellite body should be included.

\begin{figure}[ht]
\begin{center}
\noindent
  \includegraphics[width=0.48\textwidth]{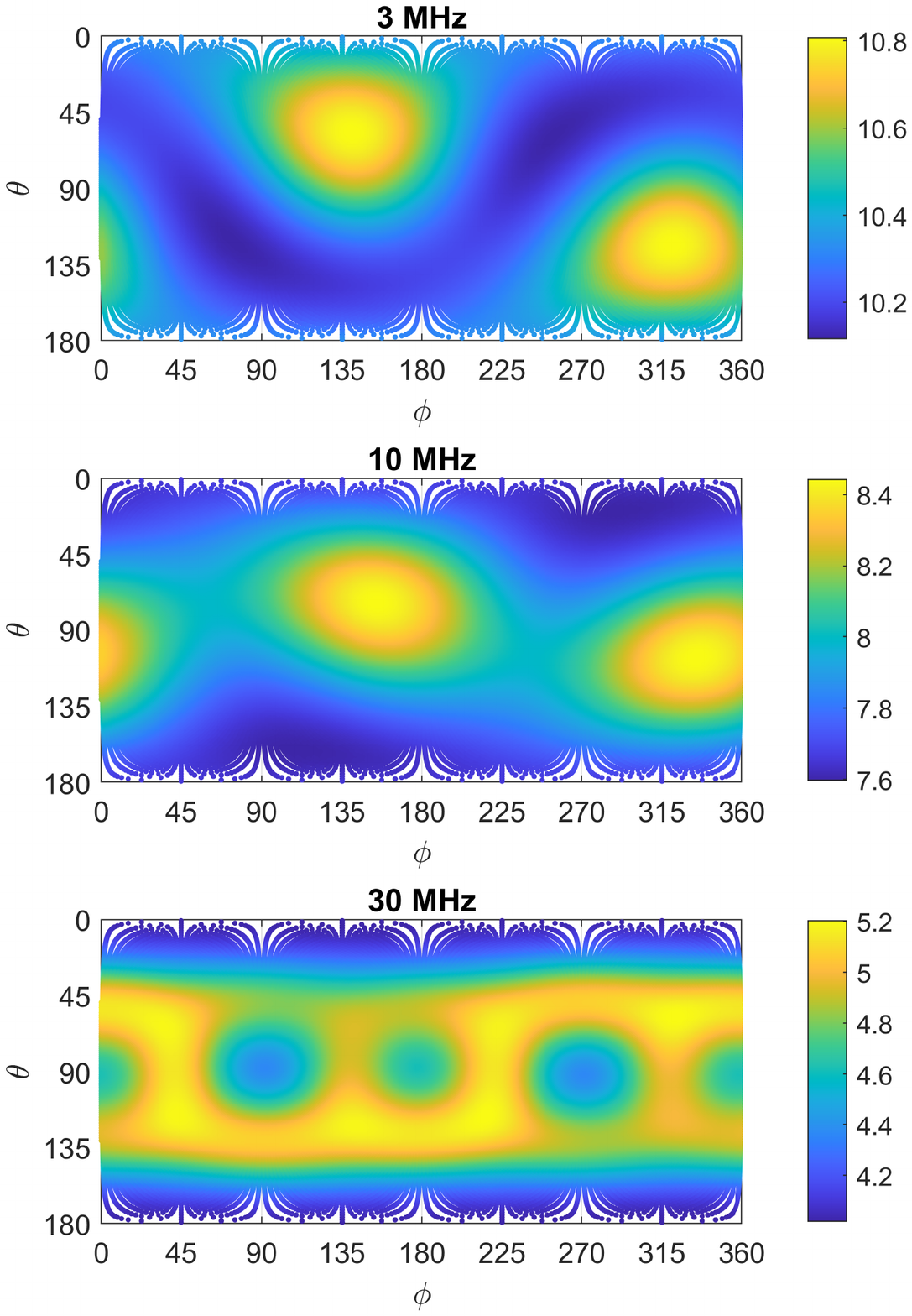}
\caption{SEFD$(\theta,\phi)$ in MJy at 3, 10 and 30~MHz for the orthogonal tripole simulated with the presence of the satellite body.}
\label{fig:SEFD_w_sat}
\end{center}
\end{figure}

Antenna temperatures $T_{{\rm ant}X}, T_{{\rm ant}Y}, T_{{\rm ant}Z}$ calculated using the simulated beam patterns and sky map in HEALPix is in good agreement with the values obtained for the analytical tripole for three frequencies. SEFD and $\Delta_{\rm SEFD}, \% = \rm (SEFD_{max} - SEFD_{min})/SEFD_{mean}$ show the most interesting outcome. From Fig.~\ref{fig:SEFD_w_sat}, we see that the presence of the metallic satellite body shifts the peak SEFD values off the $XY$-plane at 10~MHz. We see the same effect at other frequencies, and it is more noticeable the lower the frequency. The presence of the satellite body also affects the relative spread of SEFD, which increased from 1.5\% to 6.7\% at 3~{MHz}.  
We showed that Eq.~(\ref{eqn:SEFD_gen}) was successfully verified by analytical and simulated results. Simulating the tripole with the daughter satellite body showed that the SEFD($\theta, \phi$) maxima values and directions are shifted from the Galactic poles. Fig.~\ref{fig:SEFD_w_sat} shows SEFD in Jy at 3\,MHz, 10\,MHz, and 30\,MHz calculated using electromagnetic simulation of the tripole with the satellite body, as shown in Fig.~\ref{fig:tripole_w_sat}. It demonstrates that to calculate SEFD of a tripole with a realistic satellite geometry with a high accuracy, a general SEFD expression, which is given is Eq.~(\ref{eqn:SEFD_gen}), should be used. Nevertheless, for certain system engineering estimates, the short dipole approximation is ample.  

\section{Conclusion}
\label{sec:concl}
We derived and verified a system equivalent flux density, SEFD, expression valid for polarimetric radio  interferometry with multipole antennas. The expression was demonstrated using an example tripole system based on the DSL lunar orbiting satellite currently under development. The general expression shown in Eq.~\eqref{eqn:SEFD_gen} can be applied to any arbitrary multipole system and was shown to converge to the short-dipole approximation in Eq.~\eqref{eqn:SEFDs} at ultra-long wavelengths as expected. At the highest frequency of \SI{30}{\mega\hertz} considered for the DSL mission, the short dipole approximation showed small but measurable SEFD deviation. Therefore, if the highest accuracy is desired, the full expression should be used. Also, although small in terms or wavelength, the presence of the satellite body perturbs the locations of the maxima and minima of the SEFD. These effects can be studied in detail using the general SEFD expression we derived in this paper.




\section*{Acknowledgement}
\label{sec:Ack}
The authors thank of ICRAR/Curtin staff members Dr. Ben McKinley for suggesting the collaboration with NAOC, Dr. Sam McSweeney  for discussions regarding the statistical calculation, Mr. Mike Kriele for discussions regarding HEALPix processing and Mr. Daniel Ung for post-processing of FEKO results. We acknowledge key contributors to the Discovering the Sky at the Longest wavelengths project, Dr.~Fengquan Wu and Prof.~Xuelei Chen of the National Astronomical Observatories, CAS (NAOC), and Lin Wu of the National Space Science Center, CAS. We thank Yanping Cong of NAOC for supplying the sky maps. YX acknowledges the support of the Chinese Academy of Sciences (CAS) Strategic Priority Research Program XDA15020200.

\section*{Appendix: Treatment of statistics}
Following from Eq.~\eqref{eqn:varI}, we begin by considering this term
\begin{eqnarray}
\text{Var}(X_1X_2^*)&=&\left<\left|X_1X_2^*-\left<X_1X_2^*\right>\right|^2\right> \\
&=&\left<X_1X_2^*X_1^*X_2\right>-\left<X_1X_2^*\right>\left<X_1^*X_2\right> \nonumber\\
&&-\left<X_1X_2^*\right>\left<X_1^*X_2\right> +\left<X_1X_2^*\right>\left<X_1^*X_2\right>. \nonumber
\label{eqn:varX1X2}
\end{eqnarray}
The last two terms of the last line cancel, and this leaves us with
\begin{eqnarray}
\text{Var}(X_1X_2^*)= \left<X_1X_2^*X_1^*X_2\right>-\left<X_1X_2^*\right>\left<X_1^*X_2\right>. 
\label{eqn:var1a}
\end{eqnarray}
Next, we apply the formula for zero-mean joint Gaussian random variable $Z_{1,2,3,4}$~\citep{Thompson2017_rx_sys, Baudin_App3}
\begin{eqnarray}
\left<Z_1Z_2Z_3Z_4\right>&=&\left<Z_1Z_2\right>\left<Z_3Z_4\right>
+\left<Z_1Z_3\right>\left<Z_2Z_4\right> \nonumber\\
&+&\left<Z_1Z_4\right>\left<Z_2Z_3\right>,
\label{eqn:Thompson_formula}
\end{eqnarray}
to the first term in the right hand side of Eq.~\eqref{eqn:var1a}.
\begin{eqnarray}
\left<X_1X_2^*X_1^*X_2\right>&=&\left<X_1X_2^*\right>\left<X_1^*X_2\right>+\left<\left|X_1\right|^2\right>\left<\left|X_2\right|^2\right> \nonumber \\
&+&\left<X_1X_2\right>\left<X_1^*X_2^*\right>.
\label{eqn:var1c}
\end{eqnarray}
As a result,
\begin{eqnarray}
\text{Var}(X_1X_2^*)= \left<\left|X_1\right|^2\right>\left<\left|X_2\right|^2\right> 
+\left<X_1X_2\right>\left<X_1^*X_2^*\right>. 
\label{eqn:var1d}
\end{eqnarray}
The last term in the right hand side of Eq.~\eqref{eqn:var1d} contain $X_1X_2$ that are not conjugated with each other in the expectation operation $\left<.\right>$. We can write this as
\begin{eqnarray}
\left<XY\right>&=&\left<(\Re_X+j\Im_X)(\Re_Y+j\Im_Y)\right>\nonumber\\
&=&\left<\Re_X\Re_Y\right>-\left<\Im_X\Im_Y \right>+j\left<\Re_X\Im_Y\right>\nonumber\\
&+&j\left<\Re_Y\Im_X\right>.
\label{eqn:XY_iid_cplx}
\end{eqnarray}
For independent real and imaginary parts, the terms $\left<\Re_X\Im_Y\right>,\left<\Re_Y\Im_X\right>$ vanish. Furthermore, for identical  correlation in the real part and in the imaginary part, we have $\left<\Re_X\Re_Y\right>-\left<\Im_X\Im_Y \right>=0$. These are consistent with zero-mean Gaussian noise representing thermal noise. Under the foregoing conditions, $\left<X_1X_2\right>=\left<X_1^*X_2^*\right>=0$. As a result
\begin{eqnarray}
\text{Var}(X_1X_2^*)
&=&\left<|X_1|^2\right>\left<|X_2|^2\right>\nonumber \\
&=&(4k\Delta f)^2T_{sysX1}R_{X1}T_{sysX2}R_{X2}. 
\label{eqn:varsX1X2}
\end{eqnarray}
Similarly, then
\begin{eqnarray}
\text{Var}(Y_1Y_2^*)&=&\left<|Y_1|^2\right>\left<|Y_2|^2\right> \nonumber \\ &=& (4k\Delta f)^2T_{sysY1}R_{Y1}T_{sysY2}R_{Y2},\nonumber \\
\text{Var}(Z_1Z_2^*)&=& (4k\Delta f)^2T_{sysZ1}R_{Z1}T_{sysZ2}R_{Z2},\nonumber \\
\text{Var}(X_1Y_2^*)&=& (4k\Delta f)^2T_{sysX1}R_{X1}T_{sysY2}R_{Y2}. 
\label{eqn:vars}
\end{eqnarray}
we noticed the same pattern for all other cross terms, $\text{Var}(Y_1X_2^*)$,$\dots$, $\text{Var}(Z_1Y_2^*)$, which are suppressed for brevity.

Next we look at the $C$ term in Eq.~\eqref{eqn:varI}. There are $^9C_2=9!/(2!7!)=36$ terms here, but we need not examine all of them because of symmetry. There are 12 terms with $X_1$ as a leading factor
\begin{eqnarray}
C_{X1} &=& ab\text{Cov}(X_1X_2^*,Y_1Y_2^*)+ac\text{Cov}(X_1X_2^*,Z_1Z_2^*)\nonumber\\
&-&ad\left[\text{Cov}(X_1X_2^*,X_1Y_2^*)+\text{Cov}(X_1X_2^*,Y_1X_2^*)\right]  \nonumber\\
&-&ae\left[\text{Cov}(X_1X_2^*,X_1Z_2^*)+\text{Cov}(X_1X_2^*,Z_1X_2^*)\right] \nonumber \\
&+&af\left[\text{Cov}(X_1X_2^*,Z_1Y_2^*)+\text{Cov}(X_1X_2^*,Y_1Z_2^*)\right] \nonumber \\
&+&de\left[\text{Cov}(X_1Y_2^*,X_1Z_2^*)+\text{Cov}(X_1Y_2^*,Z_1X_2^*) \right] \nonumber \\
&+&d^2\text{Cov}(X_1Y_2^*,Y_1X_2^*)+e^2\text{Cov}(X_1Z_2^*,Z_1X_2^*).
\label{eqn:C_X1}
\end{eqnarray}
Now we check each term
\begin{eqnarray}
\text{Cov}(X_1X_2^*,Y_1Y_2^*) &=&\left<X_1X_2^*Y_1Y_2^*\right>-\left<X_1X_2^*\right>\left<Y_1Y_2^*\right>\nonumber\\
&=&\left<X_1X_2^*Y_1Y_2^*\right>,\nonumber\\
\text{Cov}(X_1X_2^*,Z_1Z_2^*) 
&=&\left<X_1X_2^*Z_1Z_2^*\right>, \nonumber \\
\text{Cov}(X_1X_2^*,X_1Y_2^*)&=&\left<X_1X_2^*X_1Y_2^*\right>, \nonumber \\
\text{Cov}(X_1X_2^*,Y_1X_2^*) &=&\left<X_1X_2^*Y_1X_2^*\right>,
\label{eqn:cov_check}
\end{eqnarray}
and similarly with $\text{Cov}(X_1X_2^*,X_1Z_2^*)$,$\dots$,$\text{Cov}(X_1Z_2^*,Z_1X_2^*)$. 
In Eq.~\eqref{eqn:cov_check}, terms such as $\left<X_1X_2^*\right>$, $\left<Y_1Y_2^*\right>$ vanish, assuming system noise in antenna 1 and 2 are independent and zero mean. Applying Eq.~\eqref{eqn:Thompson_formula} to Eq.~\eqref{eqn:cov_check}, we get
\begin{eqnarray}
\left<X_1X_2^*Y_1Y_2^*\right>
&=&\left<X_1Y_1\right>\left<X_2^*Y_2^*\right>, \nonumber\\
\left<X_1X_2^*Z_1Z_2^*\right>&=&\left<X_1Z_1\right>\left<X_2^*Z_2^*\right>, \nonumber\\
\left<X_1X_2^*X_1Y_2^*\right>&=& \left<X_1X_1\right>\left<X_2^*Y_2^*\right>,\nonumber \\
\left<X_1X_2^*Y_1X_2^*\right> &=& \left<X_1Y_1\right>\left<X_2^*X_2^*\right>,
\label{eqn:apply_fm1}
\end{eqnarray}
and similarly we see repeating patterns with $\left<X_1X_2^*Y_1X_2^*\right>$, $\dots$,$\left<X_1Z_2^*Z_1X_2^*\right>$. The factors with differing subscripts, for example $\left<\__1,\__2\right>$, cancel to zero because of independent zero mean noise in antennas 1, 2 as in  Eq.~\eqref{eqn:cov_check}. The factors in the expectation operator, $\left<.\right>$ that are not conjugated with respect to each other vanish as explained in Eq.~\eqref{eqn:XY_iid_cplx}.

In the final analysis $C=0$ in \eqref{eqn:varI} is well justified, which is the same conclusion as the appendix in~\cite{Sutinjo_AA2021}. However, the reasoning here is more general and depends only on the assumption consistent with thermal noise properties, that is, the real and imaginary parts of any noise source are independent and and identically distributed (the noise source may be correlated, but the real and imaginary parts must be independent). We made no assumption regarding an unpolarized sky or orthogonal antennas.

\bibliographystyle{aa}
\bibliography{Sens_array_20Jul2021.bib}

\end{document}